\documentclass{elsart}
\usepackage{graphicx}
\usepackage{amsmath}

\begin{document}

\renewcommand{\thefootnote}{\arabic{footnote}}
\setcounter{footnote}{0}

\begin{frontmatter}
\title{\LARGE\bf Statistical mechanics of random graphs}

\author[a]{Zdzis\l{}aw Burda\thanksref{apfa}},
\thanks[apfa]{Presented by Z.B. at  
{\em Applications of Physics in Financial Analysis 4}, 
Conference, Warsaw 13-15 Nov. 2003}
\author[a]{Jerzy Jurkiewicz},
\author[b]{Andr\'e Krzywicki},

\address[a]{M. Smoluchowski Institute of Physics,
Jagellonian University, Cracow, Poland}
\address[b]{ Laboratoire de Physique Th\'eorique, 
Universit\'e Paris-Sud, Orsay, France }
 
\begin{abstract}
We discuss various aspects of the statistical formulation of the theory
of random graphs, with emphasis on results obtained in a series of
our recent publications.  

\end{abstract}

\end{frontmatter}
The backbone of a generic complex system can be represented by a graph: 
the nodes refer to the system subunits and the links represent the 
interactions. Such a representation can be used, for example, to describe 
an ensemble of agents interacting in an economical framework or a
telecommunication network. Other examples of natural networks can
be found in recent reviews \cite{ab}.

In this communication we will discuss a statistical formulation of the 
graph theory \cite{bck,bk,bbjk,bjk}. The basic concept in this approach
is that of a statistical ensemble: a configuration space (set of graphs) 
is endowed with a probability measure. Instead of measuring an observable 
on a single graph one measures its average in the ensemble. One learns 
about stability and typicality of graphs by measuring fluctuations 
in the ensemble. This {\em synchronic} approach is complementary to the
diachronic one, where graphs are made up step by step, by adding
successive nodes and links.

The first step of the construction is to define the configuration space by 
choosing the set of graphs to be studied. Of course, there is much freedom in 
this choice. Here, for definiteness and unless specified otherwise, we will 
consider simple, undirected, labeled graphs with $N$ vertices and $L$ links. 
Graphs are, in general multiply connected.

It is convenient to label the nodes of a graph by an index $i=1,\dots,N$. A 
labeled graph can be represented by an adjacency matrix $A$ whose elements 
are $A_{ij} = 1$  if $i$ and $j$ are neighbors and $A_{ij}=0$ otherwise. 

The adjacency matrix of an unoriented graph with $N$ nodes and $L$ links is 
an $N\times N$ symmetric matrix with zeros on the diagonal and $L$ 
unities above (and below) the diagonal. Denote by $M_N$ the set of all 
$N\times N$ symmetric $0/1$ matrices with zeros on the diagonal and by 
$M_{NL}$ the subset where in addition the number of unities above the 
diagonal equals $L$. The partition function for the Erd\"os-R\'enyi 
\cite{er} ensemble of graphs with $N$ nodes and $L$ links can be written, 
up to some irrelevant normalization factor, as:
\begin{equation}
Z =  \sum_{A \in M_N} 
\delta\left(L - \frac{1}{2} \mbox{tr} A^2\right) 
=   \sum_{A \in M_{NL}} 1
\label{zer}
\end{equation}
Let $O=O(A)$ be an observable, a quantity defined on graphs. One is 
interested in the average over the ensemble 
\begin{equation}
\langle O \rangle = \frac{1}{Z} \sum_{A \in M_{NL}} O(A) 
\end{equation}
and in the fluctuations: $\langle O^2 \rangle - \langle O \rangle^2$ and
the higher order ones, in the large $N$ limit. The number of unities
in a row $i$ of the matrix $A$: $q_i = \sum_{j} A_{ij}$ is equal to the 
number of links emerging from the node $i$, and is called the node 
degree. The probability that a randomly chosen node of the graph has 
degree $q$ is 
\begin{equation}
p(q) = \left\langle \frac{1}{N} \sum_{i} \delta(q-q_i)\right\rangle 
\end{equation}
It is not very difficult to calculate $p(q)$ in the $N \to \infty$ limit
starting from (\ref{zer}). The result is Poissonian 
\begin{equation}
p(q) = \frac{\alpha^q}{q!} e^{-\alpha}
\end{equation}
where the constant $\alpha$ is determined by the ratio $L/N$, kept fixed 
in the limit\footnote{Notation: given a positive measure and an observable 
depending on node degrees, say $w(q)$ and $O(q)$, we write $\langle O\rangle_w$ 
for $\sum_q O(q) w(q)/\sum_q w(q)$.}: 
$\alpha \equiv \langle q\rangle_p = 2L/N$. 
However, in most interesting real networks the degree distribution is skew. 
In the so-called {\em scale-free} networks it has a fat tail extending over 
several decades and well fitted with a power law: 
$p_{\rm exp}(q) \sim q^{-\beta}$ . 
The most conservative extension of the Erd\"os-R\'enyi ensemble consists in 
introducing an additional statistical weight $W(A)$ for graphs, replacing 
the partition function (\ref{zer}) by \cite{bck,bk}:
\begin{equation}
Z =  \sum_{A \in M_{NL}} W(A)
\end{equation}
The corresponding averages read now
\begin{equation}
\langle O \rangle = \frac{1}{Z} \sum_{A \in M_{NL}} W(A) O(A) 
\end{equation}
The simplest choice for the statistical weight is:
\begin{equation}
W(A) = w(q_1) w(q_2) \dots w(q_N)
\label{uc}
\end{equation}
In the $N \to \infty$ limit one gets now
\begin{equation}
p(q) = \frac{w(q)}{q!} \exp (Aq - B)
\label{1part}
\end{equation}
where the parameters $A,B$ are chosen in such a way that 
$\sum_q p(q) = 1$ and $\langle q\rangle_p = \sum_{q} q p(q) = 2L/N$. 
Moreover, the probability that a randomly chosen graph has 
degrees $q_1,q_2,\dots,q_N$, factorizes:
\begin{equation}
p(q_1,q_2,\dots, q_N) = p(q_1) p(q_2) \dots p(q_N)
\label{npart}
\end{equation}
This is why one refers to this model as to the model of "uncorrelated networks". 
For scale-free networks the asymptotic results (\ref{1part})-(\ref{npart}) are 
only partly true because finite-size effects strongly affect the tail of the 
degree distribution: First of all, at finite $N$ the tail of $p(q)$ cannot
extend to infinity because there exists some $q_{\rm max}$
such that the expected number of nodes with $q > q_{\rm max}$
is less than unity. Neglecting correlations one finds the scaling law
\begin{equation}
q_{\rm max} \sim N^{1/(\beta-1)}
\label{cons}
\end{equation}
Furthermore, as shown in \cite{bk}, the condition that the graphs are simple,
{\em i.e.} self and multiple connections between nodes are absent, implies that
\begin{equation}
q_{\rm max} \sim N^{1/2}
\end{equation}
which for $2 < \beta < 3$ is stronger than (\ref{cons}). Hence, in this case, 
not only the degree distribution is cut but also specific correlations are 
generated at finite $N$.

By choosing the appropriate weight function $w(q)$ for uncorrelated networks  
one can reproduce the experimentally observed degree distribution, modulo
the above mentioned finite-size effects. We have constructed a numerical
algorithm, enabling one to simulate the model on a computer. We have also
obtained some further analytic results. 

In particular, the model is analytically solvable, when one restricts one's 
attention to tree graphs. In this case $\langle q\rangle_p =2$, since $L=N-1$. 
Assume that $w(q) \sim q^{-\gamma}$ at large $q$. What is the shape of the degree
distribution? The general result (\ref{1part}) no longer holds, because trees
constitute a negligible fraction of all possible graphs. It turns out that an
interesting phase structure emerges \cite{bck}:

(a) When $\langle q\rangle_w =2$ the trees are scale-free, with the degree
distribution equal to $qw(q)$, up to normalization.

(b) When $\langle q\rangle_w < 2$ one finds that the degree distribution
is up to normalization equal to $qw(q)$ for most of the range of $q$, a 
singular node showing up at $q$ of the order of $N$. The winner-takes-all 
scenario holds.

(c) When $\langle q\rangle_w > 2$ the degree distribution falls exponentially,
the scale-free input is forgotten.

Notice, that the scale-free regime is unstable with respect to small distortions
of the input weight. Further information is obtained when one calculates the 
fractal dimension $d_H$ controlling the average shortest path $r$ between a 
pair of nodes
\begin{equation}
\langle r\rangle \sim N^{1/d_H}
\end{equation}
It is known that the generic intrinsic fractal dimension of trees is $d_H=2$.
This is also what one finds in the case (c) and in the case (a) when $p(q)$ 
falls faster than $q^{-3}$, {\em i.e.} when $\beta >3$. When $2 < \beta <3$
\begin{equation}
d_H = \frac{\beta-1}{\beta-2}
\end{equation}
In the case (b) $d_H = \infty$. 

It is interesting to keep 
the same microstate weights as before, but assume that 
trees are endowed with a causal structure \cite{bbjk}. We say that this is 
the case when the node labels always appear in growing numerical order as one 
moves along the tree from the root - we have rooted trees in mind - towards an 
arbitrary node. Hence only a subclass of labelings is accepted. It turns out 
that the most popular growing network models can be reformulated in this static 
formalism. The original results are recovered in an elegant fashion. This shows 
that the widely accepted distinction between growing and equilibrium networks 
is not really correct, the two approaches are just complementary. Among new 
results is the calculation of the fractal dimension: Remarkably enough, we 
find that it is generically infinite, $d_H= \infty$, in contrast to what 
happens in maximally random trees (see above). 

Uncorrelated networks have a local tree structure. This is a well known fact 
in the context of the Erd\"os-R\'enyi theory. The same arguments hold in the 
generalized set-up. This tree structure persists when simple internode 
correlations are introduced. Actually, a general recipe generating short 
loops in static graph models was missing in the literature, until recently. 
We have succeeded to make a progress in this matter \cite{bjk,bjk2}. One
should mention that short loops are a common feature of natural networks. In
particular, the clustering coefficient is relatively large. 

The clustering coefficient for a given vertex $i$ is defined as 
$C_i = \frac{T_i}{q_i(q_i-1)/2}$, where $T_i$ is number of triangular
loops, called also three-cycles, meeting at $i$.  The clustering coefficient 
of a graph is just the average of $C_i$ over nodes. 
The reason why the coefficient is small for uncorrelated graphs
is that the number of three-cycles $T= \frac{1}{6} \mbox{tr} A^3$
is small. One can show that for the Erd\"os-R\'enyi graphs
the total number of three-cycles approaches a fixed constant, 
$\langle T \rangle = \alpha^3/6$
for $N\rightarrow \infty$. A similar result holds for cycles with a larger 
number of links. Hence, the chance of finding a cycle on a large network 
is close to zero. This is a manifestation of the local tree structure of
graphs.

We consider therefore a generalized model for graphs by adding to the 
Hamiltonian an interaction term favoring the formation of three-cycles. 
Hence, the microstate weights (\ref{uc}) are modified as 
follows\footnote{Actually, 
in ref. \cite{bjk} the unperturbed model is that of Erd\"os-R\'enyi ($W(A)=1$), 
the general case will be discussed in our forthcoming publication \cite{bjk2}.}:
\begin{equation}
W(A) \rightarrow  W(A) \exp\left( \frac{1}{6} G \mbox{tr} A^3\right)
\end{equation}
The resulting model has two phases: the crumpled and the perturbative one. 
The crumpled phase is dominated by graphs which maximize the number of 
three-cycles, which is of order $N^{3/2}$. 
For any $G>0$ and for large $N$ the term $\sim G N^{3/2}$ 
in the exponent exceeds the entropy \cite{str}.
Thus the corresponding configuration plays the role of
the ground state of the model for any $G>0$. 
The perturbative phase is obtained by letting the interaction Hamiltonian 
to act softly on the uncorrelated graphs ($G=0$). We have developed 
the corresponding perturbation theory. 
It turns out that the two phases are separated by a 
free energy barrier similar to that in a first
order phase transition. Here however the barrier has
an additional important feature. When $N\rightarrow \infty$
the barrier and the stability range of the perturbative phase increase.
This means that at large $N$ a random walker, representing a local
process in the configuration space, will never be able
to roll over the barrier and to reach the ground state. 
When the uncorrelated graphs
are those of Erd\"os-R\'enyi the value $G=G_{out}$ 
of the coupling constant where the system jumps 
to the crumpled phase scales logarithmically with 
$N$: $G_{out} = x_{out} \ln N$. By summing the leading diagrams we
get the number of three-cycles in the perturbative phase, $G<G_{out}$: 
\begin{equation}
\langle T \rangle = \frac{\alpha^3}{6} e^{G} = \frac{\alpha^3}{6} N^x
\end{equation}
where the effective coupling constant $x<x_{out}$
is defined by $G= x \ln N$. Thus, in this new theory
the number of three-cycles grows with $N$. It is a first step
towards a theory of graphs with a non-trivial clustering.

The random graph theory formulated in the language of statistical 
mechanics can be studied using the dynamical Monte-Carlo techniques 
\cite{bck,bk}. Actually, all our analytic results were checked and 
confirmed by such numerical simulations. The idea behind the Monte 
Carlo technique is to invent a Markovian process performing a sort 
of random walk in the configuration space and sampling configurations
with the frequency proportional to $W(A)$. If the process is ergodic 
and the transition probability fulfills the detailed balance
condition the process generates configurations with the required 
frequency. A good candidate for such a process is a sequence of 
rewirings performed with the Metropolis probability. Naively, to 
encode a $N\times N$ adjacency matrix one requires the quadratic 
($N^2$) storage capacity. However, since the matrix is sparse and
only the positions of $L$ it's elements are relevant, one can introduce 
a linear storage structure \cite{bck} which in practice allows one 
to code networks with up to $10^6-10^7$ nodes. Finally, let us
mention that our code not only produces graphs but also simulates
a thermal motion. This was important in refs. \cite{bjk,bjk2}.
 
\bigskip
\noindent
{\bf Acknowledgments:}
This work was partially supported by the EC IHP Grant
No. HPRN-CT-1999-000161, by the Polish State Committee for
Scientific Research (KBN) grants 
2P03B 99622 (2002-2004) and 2P03B-08225 (2003-2006),
and by EU IST Center of Excellence "COPIRA". Laboratoire de Physique
Th\'eorique is Unit\'e Mixte du CNRS UMR 8627.


\begin{thebibliography}{99}
\bibitem{ab} R. Albert, A.-L. Barabasi, Rev. Mod.
Phys. {\bf 74}, 47 (2002); 
S. N. Dorogovtsev, J.F.F. Mendes,
Adv. Phys. {\bf 51}, 1079 (2002) and
{\em Evolution of Networks: from Biological
Nets to to the Internet and WWW},
(Oxford Univerity Press, New York, 2003);
M.E.J. Newman, SIAM Review {\bf 45}, 167 (2003).
\bibitem{bck} Z. Burda, J.D. Correia, A. Krzywicki,
Phys.Rev. {\bf E64} (2001) 046118
\bibitem{bk} Z. Burda, A. Krzywicki, Phys. Rev. {\bf E67}, 046118 (2003).
\bibitem{bbjk} P. Bialas, Z. Burda, J. Jurkiewicz, A. Krzywicki, Phys. Rev. 
{\bf E67}, 066106 (2003).
\bibitem{bjk} Z. Burda, J. Jurkiewicz, A. Krzywicki, cond-mat/0310234,
to be published in Phys. Rev. E
\bibitem{er} Erd\"os-R\'enyi theory, see e.g. B. Bollob\'{a}s,
{\em Random graphs}, Academic Press, New York 1985
\bibitem{bjk2} Z. Burda, J. Jurkiewicz, A. Krzywicki, in preparation.
\bibitem{str} D. Strauss, SIAM Review {\bf 28}, 513 (1986).
\end{thebibliography}
\end{document}